# $Ca_3(Ru_{1-x}Cr_x)_2O_7$: A new paradigm for spin valves

G. Cao[1], V. Durairaj[1], S. Chikara[1], L.E. DeLong[1], and P. Schlottmann[2]

[1] Department of Physics and Astronomy, University of Kentucky,

Lexington, KY 40506, USA

[2] Physics Department, Florida State University, Tallahassee, FL 32306, USA

The spin valve effect is a quantum phenomenon so far only realized in multilayer thin films or heterostructures. Here we report a strong spin valve effect existing in *bulk* single crystals of $Ca_3(Ru_{1-x}Cr_x)_2O_7$ having an anisotropic, bilayered crystal structure. This discovery opens new avenues to understand the underlying physics of spin valves, and fully realize its potential in practical devices.

A spin valve is a device structure whose electrical resistance can be manipulated by controlling the relative spin alignment of adjacent metallic, magnetic layers separated by nonmagnetic insulating layers. Spin valves not only have technological potential as magnetic sensors and read-heads for hard drives, but also present fundamental challenges to magnetotransport, and are intensively studied in materials physics and engineering. The spin valve effect is thought to be a delicate quantum phenomenon that depends upon precision deposition and nanoscale patterning of artificial thin-film heterostructures whose quality and performance are difficult to control [1]. Here we demonstrate that a novel, strong spin valve effect exists in *bulk* single crystals of $Ca_3(Ru_{1-x}Cr_x)_2O_7$ having an anisotropic, bilayered crystal structure.

The layered ruthenates $A_{n+1}Ru_nO_{3n+1}$ (A = Ca or Sr) span almost every ordered state known in condensed matter physics, and exhibit a variety of phenomena that are among the most intriguing and challenging in contemporary solid state physics, because of the strong cross-couplings between spin, orbit and lattice degrees of freedom [2-5]. Two excellent examples are the p-wave superconductor $Sr_2RuO_4$ [2] and $Ca_3Ru_2O_7$ [3]. The latter compound has a bilayer structure (Fig.1) and undergoes an antiferromagnetic (AFM) phase transition at $T_N$ = 56 K in low fields, followed by a Mott-like metal-insulator transition at $T_{MI}$ = 48 K [3]. The AFM ground state consists of ferromagnetic (FM) bilayers stacked antiferromagnetically along the orthorhombic **c-**axis. An external magnetic field **B** leads to novel magnetoresistive properties [4-9] that are consequences of strong spin-orbit coupling, including Shubnikov-de Haas oscillations within a partially-gapped Mott state [7,8], and oscillatory magnetoresistance that is periodic in magnetic field B (not 1/B!) [10]. In particular, colossal magnetoresistance (CMR), which is *driven* by ferromagnetism in all other known materials [4], is attained in $Ca_3Ru_2O_7$ by *avoiding* a FM state [4-6], hinting at a novel mechanism based upon orbital ordering within a highly anisotropic, antiferromagnetic metallic (AFM-M) state [3-8].

$Ca_3Ru_2O_7$ is then a truly unique material for further study. Recent density functional calculations suggested that it is nearly "half-metallic" (the density of states of minority spins is gapped) with a potential for bulk spin valve behavior [11]. However, this prediction has not been experimentally confirmed, simply because the uniform magnetic exchange coupling within the Ru-O bilayers of pure $Ca_3Ru_2O_7$ does not render the contrasting magnetic coercivities (i.e., "soft" and "hard" layers) necessary for spin valve behavior. Moreover, the AFM-M state exists only over a narrow, 8 K interval between $T_{MI}$ and $T_N$ [3], which is too restrictive for the underlying physics to be fully investigated.

We have pursued transport and thermodynamic studies of single-crystal $Ca_3(Ru_{1-x}Cr_x)_2O_7$ with $0 \leq x \leq 0.20$ [12], since Cr substitution is uniquely effective for widening the stability range of the AFM-M state while preserving the most intriguing properties of pure $Ca_3Ru_2O_7$. This is particularly evident in the magnetization data for $x = 0.20$ and **B** ∥ **a**-axis ($M_a$), which clearly show that Cr substitution lowers $T_{MI}$ and rapidly raises $T_N$, greatly extending the AFM-M state from an 8 K interval to one over 70 K (Fig. 2a). The magnetization for **B** ∥ **b**-axis ($M_b$) exhibits the same increase in $T_N$, but with an upturn below $T_{MI}$ for $x > 0$, suggesting that spin canting is present in the AFM-M state for **B** ∥ **b**-axis (Fig. 2b). The decrease of $T_{MI}$ implies delocalization of d-electrons and the increase in $T_N$ signals an enhanced exchange coupling between nearest-neighbor spins. This is similar to $SrRu_{1-x}Cr_xO_3$ and $CaRu_{1-x}Cr_xO_3$ where the FM state is either strongly enhanced [14-16] or induced [17] by Cr doping, respectively.

We now focus on probing the AFM-M state of a representative composition, x = 0.17, as a function of temperature, magnetic field strength and orientation. The magnetizations $M_a$ and $M_b$ of $CaRu_{0.83}Cr_{0.17}O_3$ indicate $T_{MI}$ = 42 K and $T_N$ = 86 K at B = 0.5 T, as shown in Fig. 2c. The specific heat C(T) exhibits two mean-field-like 2$^{nd}$ order phase transitions at $T_{MI}$ and $T_N$ in C/T (Fig. 2c, right scale). Both transitions are slightly smeared as a consequence of an inhomogeneous Cr distribution. The transition anomaly at $T_{MI}$ is sharper with an approximate jump $\Delta C \sim 0.31$ R (R = 8.314 J/mole K), and the other one near T = 83 K (slightly lower than $T_N$ = 86 K) is broader with a smaller $\Delta C \sim 0.28$ R. A fit of the low-T data to $C = \gamma T + \beta T^3$ for 1.7 K < T < 30 K yields coefficients of the electronic and phonon contributions, $\gamma \sim 31$ mJ/mol K and $\beta \sim 3.0 \times 10^{-4}$ mJ/mol K$^3$, respectively, which are comparable to the values at x = 0 [18, 19]. A similar fit to the data immediately above and below $T_N$ yields the estimates $\gamma$ = 680 and 79 mJ/mole K$^2$ for the paramagnetic and AFM-M states, respectively. The large difference between the $\gamma$ coefficients of the two states implies that a substantial reduction of Fermi surface accompanies the onset of AFM. The measured transition entropy is approximately 0.037 R, which is much smaller than 2Rln3 expected for complete ordering of localized S = 1 spins. On the other hand, AFM ordering among itinerant spins (e.g., as in a spin-density wave), should produce $\Delta S \sim \Delta\gamma T_N$ and a mean-field-like step $\Delta C \sim (1.43)\Delta\gamma T_N$, which is qualitatively consistent with our data, but implies $\Delta\gamma/R \sim 0.0024$ K$^{-1}$. Alternatively, if the transition at $T_{MI}$ were due to the formation of a charge density wave, then the expected entropy change $\Delta S \sim \Delta\gamma T_{MI}$ yields $\Delta\gamma/R \sim 8.8 \times 10^{-4}$ K$^{-1}$, which is in still greater disagreement with our data. The measured value of $\Delta\gamma/R$ therefore suggests that the anomaly at $T_N$ is inconsistent with both conventional itinerant and localized pictures of AFM [19], and thus indicates a more complex spin ordering within the AFM-M state.

Indeed, $CaRu_{0.83}Cr_{0.17}O_3$ is driven toward a FM state at B > 5 T for **B || a**-axis, as $T_N$ decreases at a rate of $\Delta T_N/\Delta B$ = -1.4 K/T. However, $T_N$ is readily suppressed at an astonishing rate $\Delta T_N/\Delta B$ = -7.5 K/T for **B || b**-axis, suggesting that the magnetic lattice softens. Such an unusual magnetic anisotropy is also reflected in the **c**-axis resistivity $\rho_c$ at B = 0, which sharply drops below $T_N$ (Fig.2d), despite the Fermi surface reduction indicated by the reduction of $\gamma$ at $T_N$. A possible explanation could be the reduction of spin-scattering with increasing magnetic order. This behavior differs from that of other AFM-M systems such as Cr and Mn [20, 21] where the onset of AFM order at $T_N$ is accompanied by a rise in resistivity. Moreover, the conductivity in the AFM-M state is strongly spin-dependent and extremely anisotropic: For **B || a**-axis, the AFM state becomes semiconducting despite the emerging field-induced FM state for B > 5 T; conversely, for **B || b**-axis, $\rho_c$ *decreases* with applied field, as shown in Fig. 2d. All in all, in the range $T_{MI}$ < T < $T_N$ the system remains an AFM-M for **B || b**-axis, but becomes FM and semiconducting for **B || a**-axis. This behavior indeed bears a resemblance to that predicated for half-metallic systems [22].

The field dependences of the magnetoresistivity ratio $[\rho_c(B)-\rho_c(0)]/\rho_c(0)$ and magnetization M(B,T) exhibit two important trends: **(1)** Figs. 3a and 3b reveal a metamagnetic transition that occurs at T < 30 K for $M_a$ but at T > 30 K for $M_b$, respectively; the reversed anisotropy implies that the magnetic easy axis rotates from the **a**-axis to the **b**-axis for T > 30 K. **(2)** Although $M_b$ is notably *smaller* than $M_a$ at T ≤ 30 K, $[\rho_c(7T)-\rho_c(0)]/\rho_c(0)$ for **B || b**-axis is *greater* in magnitude than that for **B || a**-axis

(Figs. 3c and 3d). These data imply an *inverse relation* exists between M (B) and [$\rho_c$(B)-$\rho_c$(0)]/$\rho_c$(0), and confirms the existence of a novel magnetotransport mechanism based on orbital order rather than a spin-polarized state [4,5].

The phenomenon centrally important to our study is that $\rho_c$(B) for **B** || **b**-axis *peaks* at a critical field $B_{C2}$ that marks a sharp change in slope of $M_b$(B). $M_b$ maximizes in the range 35 < T < 45 K via two distinct transitions, $B_{C1}$ > $B_{C2}$, culminating at a saturation magnetization $M_s$ ~3 $\mu_B$/f.u at 7 T, as shown in Figs. 3b and 4b. A rapid increase in magnetization normally implies a monotonic reduction of spin scattering with increasing field, as for x = 0 for **B** || **b**-axis (see Fig. 4a), x = 0.17 for **B** || **a**-axis (Figs. 3a and 3c), and most other materials. This expectation clearly disagrees with the peak in $\rho_c$(B) (Figs. 3d, 4b and 4c), which is observed for **B** || **b**-axis over an extended range 35 < T < 65 K.

These compelling trends can be explained within a spin-valve scenario, as sketched in Fig. 4b. The AFM inter-bilayer coupling is far weaker than the FM interaction within a bilayer for x = 0. The existence of a robust spin valve effect demands that Cr substitution result in a Ru-O layer being replaced by a Cr-rich layer in some bilayers; for x = 0.17, the replacement of a Ru-O by a Cr-O layer is likely to occur every 2 or 3 bilayers. Extremely weak superlattice peaks expected in single crystal x-ray diffraction for a highly ordered stacking of Cr-O layers have not been observed, but an ordered stacking of the Cr-O layers is not necessary for our discussion. The strong anomalies seen in C(T) along with the drastically enhanced $T_N$ are consistent with a non-random Cr distribution.

The presence of the Cr-O layer causes a spin canting at low fields and temperatures, which gives rise to the upturn in $M_b$ at $T_{MI}$ (Fig. 2). The magnetization of each un-substituted Ru-O bilayer, or **hard magnetic bilayer**, is pinned due to the strong exchange coupling within the bilayer, whereas the magnetization of a substituted Cr-rich layer, which constitutes a *soft magnetic bilayer*, more freely rotates toward the applied field because of the interrupted or weakened exchange coupling of the Cr-rich bilayers as compared to the undoped Ru-O bilayers. *Antiparallel alignment* in the soft bilayer is achieved when the spin in the Cr-rich layer fully switches at $B = B_{C2}$ (Fig. 4b). This spin switching enhances $M_b$ at $B_{C2}$ but, at the same time, changes the density of states for the up- and down-spin electrons at the Fermi surface. The probabilities of scattering for both up- and down-spin electrons are enhanced within the soft bilayers having antiparallel alignment, which requires transport electrons to flip spin to find an empty energy state, which explains the pronounced peak in $[\rho_c(B)-\rho_c(0)]/\rho_c(0)$ (Figs. 4b and 4c). The remaining antiparallel spins of Ru-O layers in both the soft and hard bilayers also switch with further increase of B, finally completing the spin alignment at $B = B_{C1}$. Scattering is now zero for the up-spin electrons and still finite for down-spin electrons. Since conduction occurs in parallel for the two spin channels, the total resistivity is chiefly determined by highly conductive up-spin electrons and, consequently, $\rho_c$ drops dramatically by as much as 40%. This magnetoresistive effect is much larger than that seen in thin-film mulilayers [1], and could have *important technological implications*.

While $B_{C1}$ decreases with T, $B_{C2}$ increases slightly with T and disappears at T > 45 K. This may be due to a reduction in the difference in soft and hard layer coercivities that becomes insignificant in applied fields at higher temperatures. Consequently, switching may occur almost simultaneously for both the Cr-rich and Ru-O bilayers, resulting in one sharp transition at $B_{C1}$ seen in both $M_b$ and $\rho_c$ (Fig. 4c), which persists up to 70 K. The B-T phase diagram given in Fig. 5 shows that all transitions, $T_N$, $T_{MI}$, $B_{C1}$ and $B_{C2}$, meet at a tetracritical point at B = 3.8 T and T = 45 K. It is in the vicinity of this point that prominent spin-valve behavior exists.

In summary, this work demonstrates the occurrence of spin-valve behavior in a ***bulk material***. Previous to this work, the spin-valve phenomenon was realized solely in heterostuctures and multilayer films [1]. Cr substitution apparently alters the density of states in the AFM-M state and creates soft and hard bilayers having antiparallel spin alignments that induce spin-valve behavior for **B** ∥ **b**-axis. Additional studies are underway to identify the exact nature of the magnetic microstructure that governs the striking magnetoresistive behavior of this system.

**Acknowledgements:** This work was supported by NSF through grants DMR-0240813 and 0552267. P.S. is supported by DoE Grant #DE-FG02-98ER45707. L.D. is supported by DoE Grant #DE-FG02-97ER45653.

single crystals are of high quality. The highly anisotropic physical properties of $Ca_3(Ru_{1-x}Cr_x)_2O_7$, another indication of the crystal quality, are used to determine the magnetic easy axis and to identify twinned crystals.

**Captions:**

**Fig. 1.** (a) The bilayer structure of $Ca_3Ru_2O_7$ in the bc-plane, and (b) the corresponding TEM image. The dark stripes are magnetic Ru-O layers and light gray stripes are insulating Ca-O layers.

**Fig. 2.** The magnetization M as a function of temperature T for (a) the **a**-axis $M_a$, (b) **b**-axis $M_b$ at B = 0.5 T for various Cr content x, (c) $M_a$ and $M_b$ at B = 0.5 T for x=0.17; right scale: C/T vs. T for x = 0.17. (d) Temperature dependence of the **c**-axis resistivity $\rho_c$ for x = 0.17 for **B** || **a**-axis and **B** || **b**-axis at B = 0 and 7 T. Note that the shaded area is the range of the AFM-M state.

**Fig. 3.** Field dependence of M for **(a)** $M_a$ and **(b)** $M_b$, and field dependence of **(c)** $[\rho_c(B)-\rho_c(0)]/\rho_c(0)$ for **B** || **a**-axis and **(d) B** || **b**-axis for $2 \leq T \leq 85$ K.

**Fig. 4.** Field dependence of $[\rho_c(B)-\rho_c(0)]/\rho_c(0)$ and M (right scale) for **B** || **b**-axis at **(a)** T = 43 K for x = 0, **(b)** 40 K for x = 0.17, and **(c)** 45 K for x = 0.17. The thin (thick)-line arrow indicates the spin direction in the Cr-O rich (RuO) layer.

**Fig. 5.** T-B diagram for x = 0.17 based on the data or **B** || **b**-axis. The thin (thick)-line arrow indicates the spin direction in the Cr-O rich (RuO) layer.

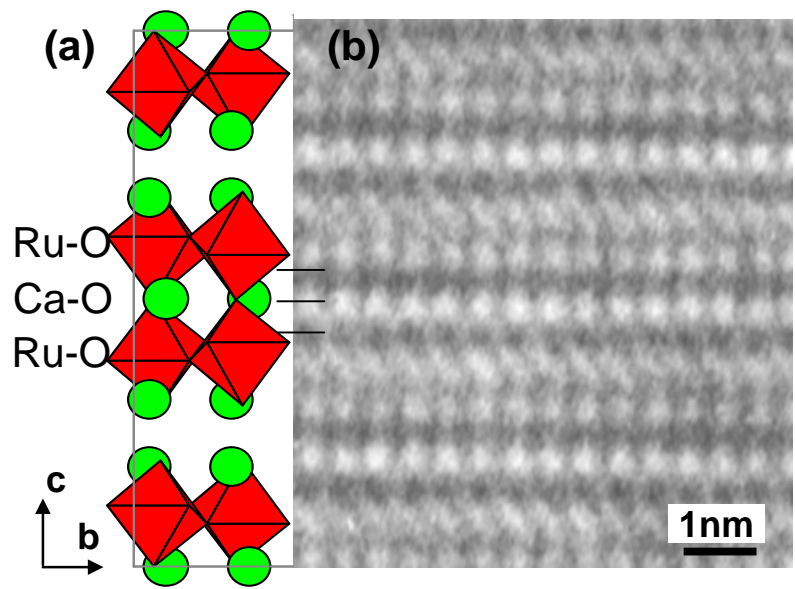

**Fig. 1**

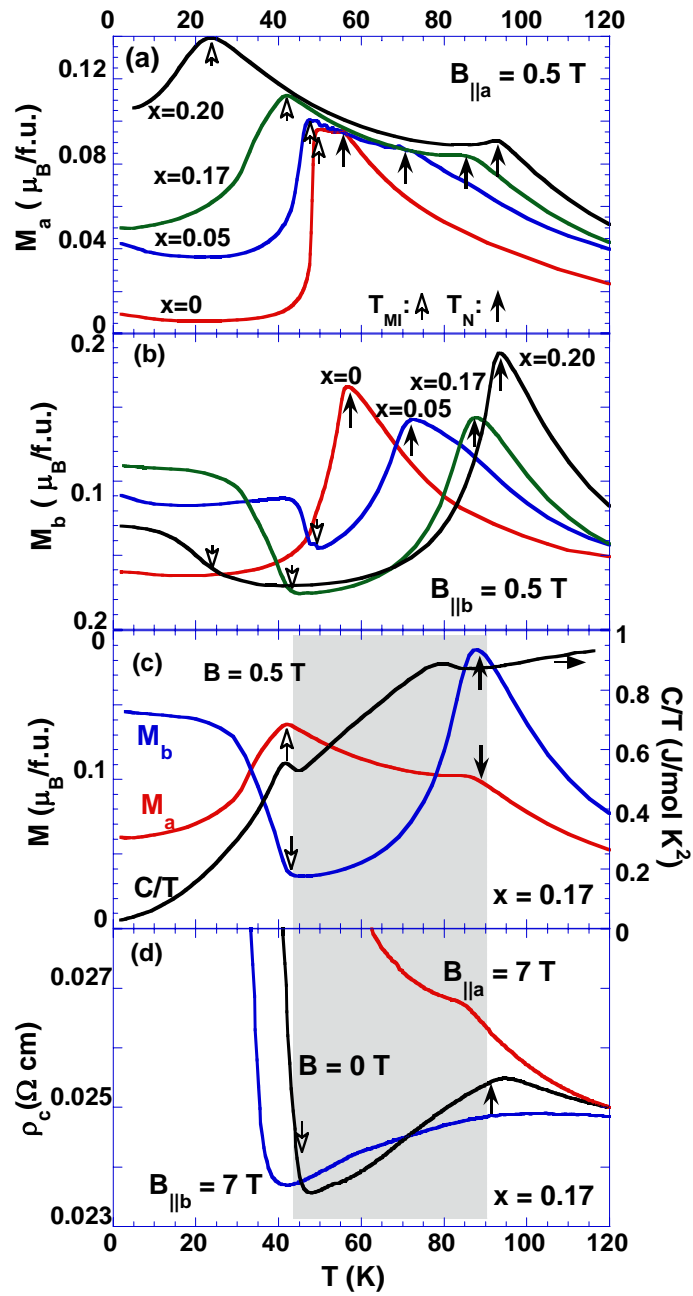

Fig. 2

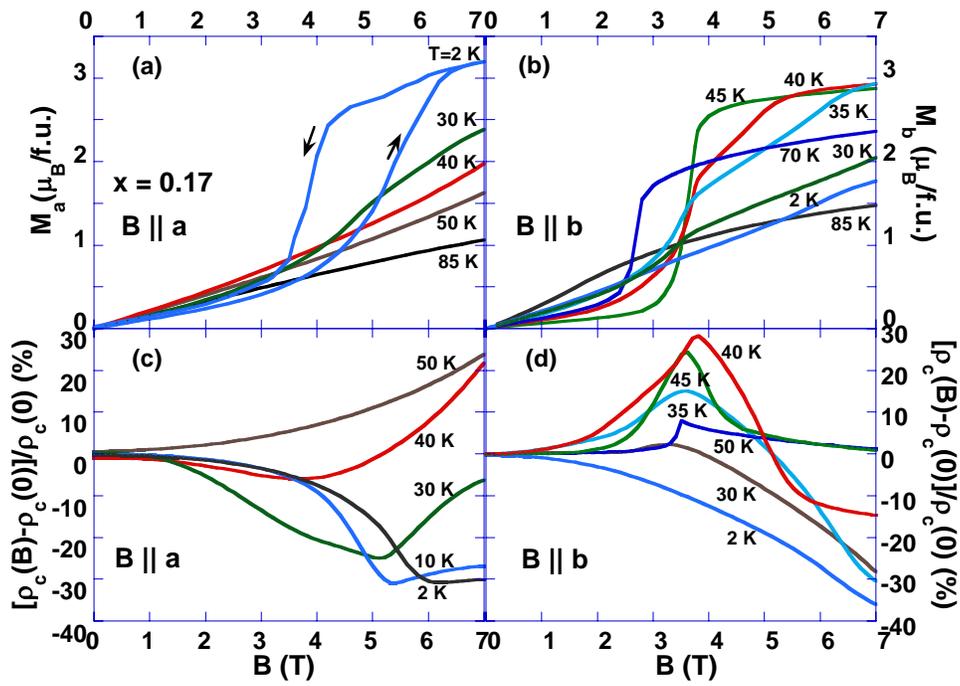

Fig. 3

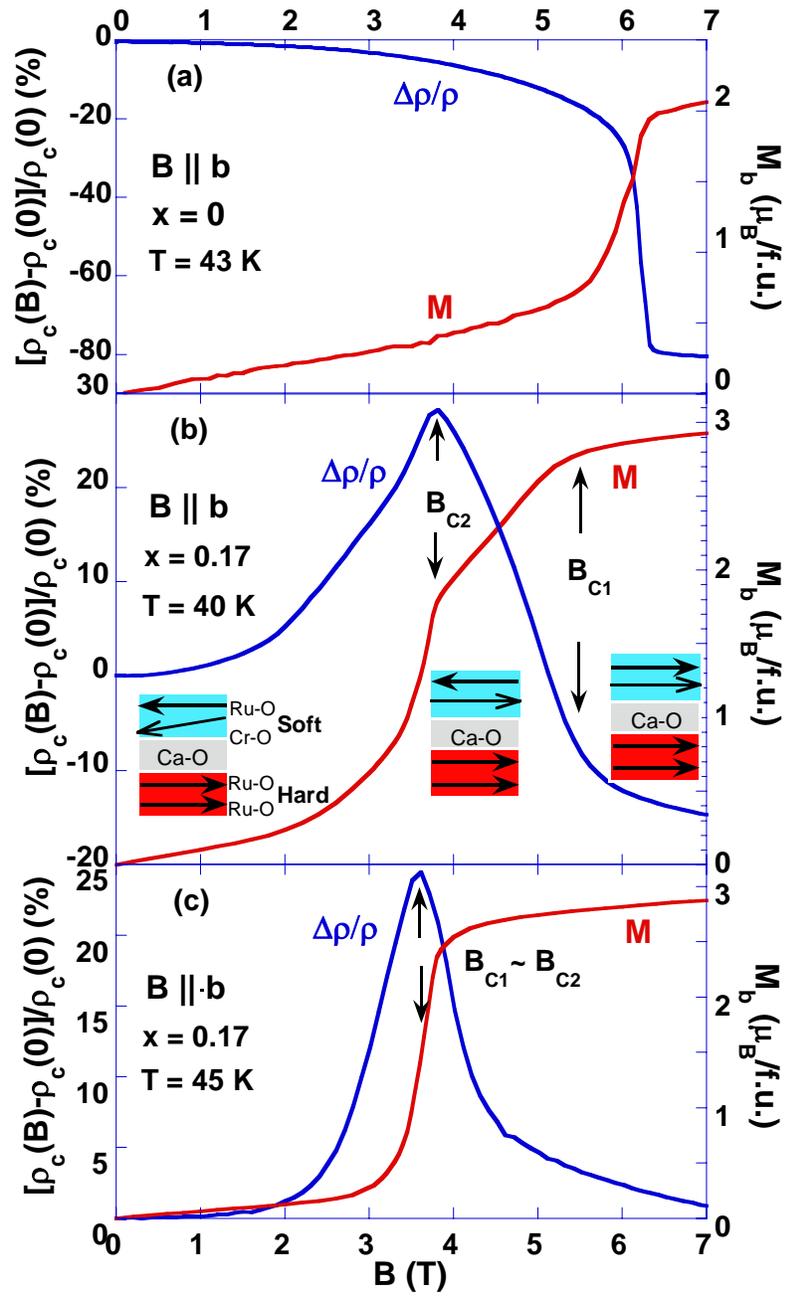

Fig. 4

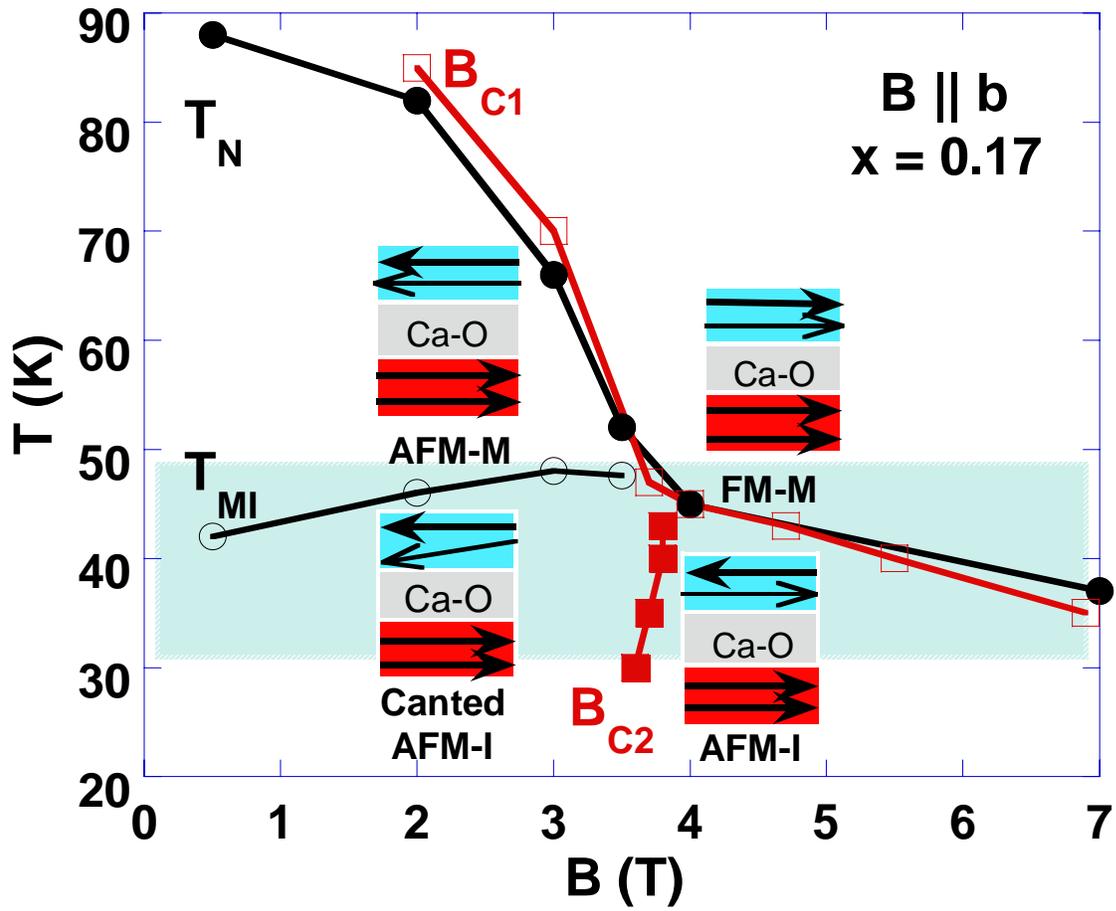

Fig. 5